\documentclass[12pt,a4paper]{article}
\usepackage{amsmath}
\usepackage{graphicx}
\begin{document}
\vspace*{5mm}
\begin{flushright}
 RRC \hspace*{13mm}\strut\\"Kurchatov Institute"\\
 IAE-6342/2 \hspace*{5mm}\strut
\end{flushright}
\vspace*{15mm}
\begin{center}
  \Large \bf S-wave  $\pi\pi$ phase shifts and scattering lengths.\\
   The model-independent analysis.
\end{center}
\vspace*{1mm}
\begin{center}
   \large {V.N.Ma\u{\i}orov,  O.O.Patarakin}
 \end{center}
\vspace*{1mm}
\begin{center}
\it  Russian  Research  Center "Kurchatov  Institute",\\
     Institute  of General  and  Nuclear  Physics,\\
     pl. Kurchatova 1, 123182 Moscow, Russian Federation\\
     E-mail: mvn@kiae.ru
 \end{center}
\begin{abstract}
The model-independent analysis of the  S- and P-wave $\pi\pi$ phase shifts
was carried out. This analysis was based on the using of the Roy equations
only and all  available experimental data from the threshold
up to dipion mass $m_{\pi\pi}=~1~GeV$.
As the results S-wave lengths were calculated: $a^0_0=(0.212\pm 0.015)\,m_{\pi}^{-1}$;\
$a^2_0=(-0.043\pm 0.010)m_{\pi}^{-1}$.
The result obtained obviously confirm the standard ChPT version.
Moreover,  additional arguments were found in favor of the
ratio of the S-wave  phase shifts $\delta^0_0(s)$ and $\delta^2_0(s)$
being independent of energy from the threshold up to  $m_{\pi\pi}=~900~MeV$.
The proportionality coefficient between the phase shifts $\eta$
is equal to $-4.66\pm0.05$.
\end{abstract}
\vspace*{3mm}
\hspace*{26mm}  Perturbation  Theory\\[10mm]
\newpage
\section{Introduction}
An investigation of the near-threshold parameters of the  $\pi\pi$
interaction  has acquired  a special role due to emergence of  QCD
theories with a broken down chiral symmetry.
During last years,  Chiral Perturbation Theory (ChPT)~\cite{Gas,Bij}
and Generalised Chiral Perturbation Theory (GChPT)~\cite{Kne} were developed.
Both these theories can describe the strong interactions at low energy.
The determinative factor in these theories is the existence of vacuum condensates violating
chiral symmetry. These theories having the same form of the effective  Lagrangian  differ
from each other by  value of  quark condensate and light quark masses.
The fact determining the choice of the version is that
the S-wave $\pi\pi$  scattering lengths $a^0_0$ and $a^2_0$ are very sensitive to the
parameters of the model and consequently are the key parameters for unambiguous
determination of the scenario of chiral symmetry violation. In this way, ChPT  predicts
the value $a^0_0$=0.220 and GChPT  $a^2_0=0.263$\footnote{The S-wave scattering lengths $a^0_0$
and $a^2_0$ are given in $m_{\pi}^{-1}$}.
So, a reliable determination of the  $\pi\pi$ lengths enables one to estimate the amount
of chiral symmetry violation and to choose thereby an adequate version of the theory.\\
During some time, despite of large accumulated experimental material on scattering lengths,
this choice has been  difficult to be made. The matter is that the experiment $K_{e4}$~\cite{Ros}
gave evidence in favor of GChPT, whereas most $\pi N\longrightarrow\pi\pi N$ experiments
inclined rather to ChPT.\\
The aim of our program, begun in~\cite{Pat} and continued in~\cite{Mai,Mai2}, was to choose
a true chiral version without using additional constraints based on chiral theories.
Therefore our calculation were based on the Roy equations only and all  available
experimental S- and P-wave phase shifts.
In our work~\cite{Mai} very large uncertainties of the $\pi\pi$ lengths were obtained
that  prevented making unambiguous choice.
In the next paper~\cite{Mai2} the additional relation linking the S-wave phase shifts was used.
This relation was received on the basis of the analysis of the S-wave behavior  above the
threshold only. Theses of chiral theories were not used at all.
As the result, the accuracy of determination of S-wave lengths $a^0_0$ and $a^2_0$
was considerably improved by means of eliminating the correlation between them.
The obtained lengths were in a good accordance with the standard ChPT version.\\
In the present work it will be shown that adding of  the new data from  the latest
$K_{e4}$ E865~\cite{Pis} experiment  to the base experimental data set used in~\cite{Mai,Mai2},
makes it possible to improve considerably the accuracy
of determination of $a^0_0$ and $a^2_0$ and for certain  to choose, without using
additional constraints, the scenario of chiral symmetry violation.
%
\section{Roy  equations}
%
The using of the general principles of unitarity, analyticity and
crossing symmetry is one of the seminal approaches to study
$\pi\pi$ interaction.
For $\pi\pi$ amplitudes, the integral equations known as "the Roy
equations" [9-11] proved to be rather useful on this path.
These equations determine the real parts of the partial wave
amplitudes which satisfy the analyticity and crossing symmetry
conditions in the $-4<s<60$ range in terms of $\pi\pi$ amplitude in
the physical $4<s<\infty$\footnote{Here and below, s is the Mandelstam variable,
$s=m^2_{\pi\pi}/m^2_{\pi}$} region.
The Roy equations combined with the unitarity relations constitute
a system of non-linear singular integral equations.
In deriving these equations, the dispersion relations with two
subtractions at fixed four-momentum transfer \textbf{t} and the crossing symmetry property
of the scattering amplitudes were used.
In present work the Roy equations were solved to get S-wave
$\pi\pi$ lengths.
All available experimental S- and P-wave phase shifts from the
threshold up to $m_{\pi\pi}=~1~GeV$ were used. And what's more the
new high-accuracy data from latest $K_{e4}$ E865~\cite{Pis}
experiment were added to the base experimental data set [12-23]
used in~\cite{Mai}.\\
For the $S_0$ wave description the phase shifts $\delta^0_0$
obtained in the $\pi N\longrightarrow\pi\pi N$ and $\pi
N\longrightarrow\pi\pi\Delta$ processes [12-17] were adopted.
From~\cite{Kam} the values of the "down-flat" set was used
only.
In the region being studied, the "down-steep" solution coincides
with "down-flat" one. Whereas the "up-flat" and "up-steep" versions
cannot be described by a smooth curve and are strongly differed
from the other data used.
The results of the $K_{e4}$~\cite{Ros,Pis} experiments were used
also.\\
For the $S_2$ wave description the phase shifts $\delta^2_0$
obtained in the $\pi^-p\longrightarrow\pi^-\pi^-\Delta^{++}$
[18-22] and $\pi^+p\longrightarrow\pi^+\pi^+n$~\cite{Hoog}
processes were adopted.
Precise values of the cross sections $\sigma_{\pi\pi}(s)$ near the
threshold were obtained in~\cite{Kerm}.
This permitted to estimate the values $\delta^2_0$ in this region
under the assumption that phase shifts $\delta^0_0$ are known.\\[1mm]
As the result the phase shifts $\delta^2_0$ and their uncertainties
were calculated by using cross sections $\sigma_{\pi\pi}(s)$ and
the values of $\delta^0_0$ near the threshold from~\cite{Ros,Pis,Alek}.
The resulting values are presented in the Table 1 (Appendix A). \
For the P-wave describing the results obtained in the
$\pi^+\pi^-\longrightarrow\pi^+\pi^-$ and
$\pi^{\pm}\pi^0\longrightarrow\pi^{\pm}\pi^0$ channels [12,14-16] were used. \
 For the case of the charged pions, the Roy equations are given by:
\begin {equation}
 \label{f1}
  Ref^I_l(s)=\lambda^I_l(s)+\frac{1}{\pi}\,\int_4^{51}\Psi^I_l(x,s)\,dx+\varphi^I_l(s),
\end {equation}
where  $\Psi^I_l(x,s)=Im\, f^0_0(x)K^I_{1l}(x,s)+Im\, f^1_1(x)K^I_{2l}(x,s)+
Im\, f^2_0(x)K^I_{3l}(x,s).$
Explicit expressions for the kernels $ K^I_{jl}(x,s)$ are given in Appendix B.
The corrections $\varphi^I_l (s)$ estimating the contributions from
the higher waves ($l\ge2 $) and from the large mass region were
adopted from~\cite{Pen}.
\begin {equation}
\begin {split}
  &\varphi^0_0(s)=13\times10^{-5}(s^2-16)\pm\Delta\varphi^0_0;\ \
  \Delta\varphi^0_0\ =\ 5\times10^{-5}(s^2-16)\\
  &\varphi^2_0(s)=13\times10^{-5}s(s-4)\pm\Delta\varphi^2_0; \quad
  \Delta\varphi^2_0\ =\ 6\times10^{-5}s(s-4)
\label{f2}
\end {split}
\end {equation}
The subtraction terms $\lambda^I_0(s)$ are expressed in terms of
the scattering lengths:
\begin {equation}
 \label{f3}
  \lambda^0_0(s)=a^0_0+\frac{s-4}{12}\,(2a^0_0-5a^2_0); \qquad
  \lambda^2_0(s)=a^2_0-\frac{s-4}{24}\,(2a^0_0-5a^2_0)   
\end {equation}
We realized  the same numerical method  to solve  the Roy equations  as in~\cite{Mai}
 without using iterative procedures.
Due to this approach the problem of convergence of the solutions
was eliminated automatically and the process of calculation of
scattering lengths $a^0_0 $ and $a^2_0 $ became absolutely
clear.
The solution of the Roy equations~(\ref{f1}) comprised some steps.
First, we performed fitting for each phase shift $\delta^I_l $ and obtained
smooth curves adequately describing
experimental data. In particular for the S-wave phase shifts
expansion~(\ref{f4}) was used:
\begin {equation}
 \label{f4}
  \delta^I_0(s)=\frac{2}{\sqrt{s}}\,(C^I_1\,q+C^I_2\,q^3+\cdots+C^I_m\,q^{2m-1})
\end {equation}
where $q =\frac12\sqrt {s-4} $ - is c.m. pion momentum and $C^I_k$
- are free parameters (I=0,2; $k=1\div4) $.
We used m=4 because the increase of the number of terms of the
series would not improve the accuracy of the smoothing.
For the $S_0$ wave, when $n=106$ experimental points were used, it
was obtained: m=4,\ $\chi^2$=137.76; \ \ m=5, \ $\chi^2$=137.28.
For the $S_2$ wave, when $n=28$: m=4,\ $\chi^2$=36.48; \ \ m=5,\ $\chi^2$=36.36.
Thus, the describing the S-wave phase shifts by means of the used
polynomial is stable.
Experimental values of phase shifts and fitting curves~(\ref{f4})
%
 \begin{figure}[t]
   \includegraphics[width=\textwidth, height=4in]{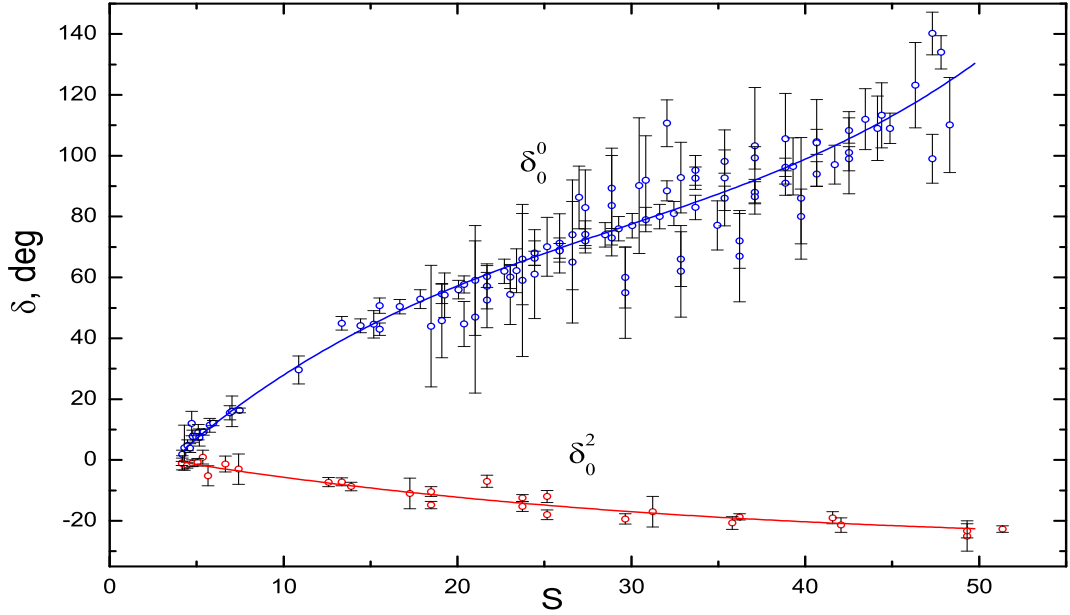}
   \caption{S-wave  $\pi\pi$   phase  shifts.  The solid curves  represent the result
                   of  fitting in  terms  of  expression~(4).}
  \end{figure}
%
are shown in Fig.1.
In the present study, we assume, as in~\cite{Mai}, that in the
energy range considered the P-wave is determined by the
$rho$-resonance almost completely.\\
On the second stage, the obtained smooth dependencies
$\delta^I_l(s)$ were used as input for the Roy equations~(\ref{f1})
and after integration, the subtraction terms $\lambda^I_l(s)$ were
calculated.\\[1mm]
The values of $Ref^I_0(s_i)$ were taken at each experimental point
$s_i$ where the phase shifts $\delta^I_0(s_i)$ were
measured.
By solving the Roy equations for each values of $s_i$, we obtained
the values of the subtraction terms $\delta^I_0(s_i)$ and their
statistical errors $\sigma_{\lambda^I_0}(s)$ from experimental data
on the $\pi\pi$ phase shifts.
This errors are determined ultimately by the errors of the phase
shifts $\delta^I_0(s_i)$ and were calculated by means of the
standard rule of propagation of errors.
It should be emphasized that the expression for the uncertainties
$\sigma_{\lambda^I_0}(s)$ does not contain the theoretical errors
 $\Delta\varphi^I_0(s)$, since they are not, generally speaking, statistical:
they change the behavior of the function $\varphi^I_0(s)$ simultaneously for
all s.
Because of this, the theoretical corrections $\varphi^I_0(s)$
behave as random functions with respect to
$\Delta\varphi^I_0(s)$.
Therefore contribution of the uncertainties $\Delta\varphi^I_0(s)$
in the errors of the lengths $a^0_0$ and $a^2_0$ should be
calculated separately.\\
At the conclusion stage we carried out fitting of the dependencies
$\lambda^I_l(s_i)$ using terms~(\ref{f3}) and determined the S-wave
 $\pi\pi$ lengths.
Such approach enabled us to study in detail each isotopic channel of
the Roy equations by evaluating the contribution of each phase
shift $\delta^I_l$ in the resulting subtraction terms
$\lambda^I_0(s)$ separately.
It is in this way, it was found that the phase shifts $\delta^2_0$,
obtained in the " electronic experiment"~\cite{Hoog},
lead to the result which contradicts  considerably  the result obtained by
processing the rest of the phase shifts $\delta^2_0$ data base.
Therefore, we did not use the phase shifts from~\cite{Hoog} in the
present study. This problem will be considered below.
%
 \begin{figure}[t]
   \includegraphics[width=\textwidth, height=4.5in]{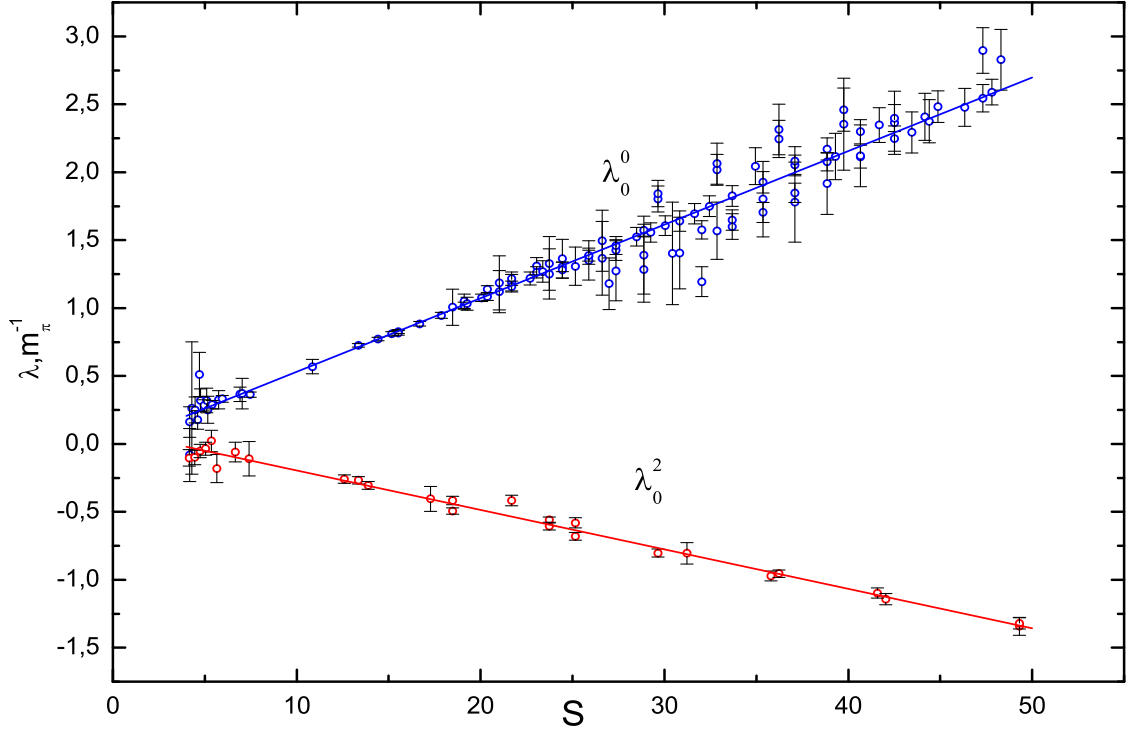}
   \caption{Subtraction terms $\lambda^0_0(s)$ and $\lambda^2_0(s)$. The straight  lines
                  represent the result  of  fitting in terms of expression~(3).}
 \end{figure}
%
Now the solutions of the Roy equations obtained for each isotopic
channel will be given. Hereinafter r is a factor of correlation
between $a^0_0$ and $a^2_0$.
In the isotopic channel I=0, it was obtained by fitting the
subtraction term $\lambda^0_0(s_i)$:
\begin {equation}
 \label{f5}
    a^0_0=0.207\pm0.009; \quad   a^2_0=-0.047\pm0.005; \quad   r=0.989
\end {equation}
$\chi^2$/NDF=127/106.
After taking into consideration the theoretical error
 $\Delta\varphi^0_0$ \, it was received:
\begin {equation}
 \label{f6}
    a^0_0=0.207\pm0.015; \quad   a^2_0=-0.047\pm0.011; \quad   r=0.980
\end {equation}
In the isotopic channel I=2, it was got by fitting the subtraction
term $\lambda^2_0(s_i)$:
\begin {equation}
 \label{f7}
         a^0_0=0.295\pm0.042; \quad   a^2_0=-0.022\pm0.014; \quad   r=0.996
\end {equation}
$\chi^2$/NDF=32.6/25.
After taking into consideration the theoretical error
$\Delta\varphi^2_0$, it was obtained:
\begin {equation}
 \label{f8}
         a^0_0=0.295\pm0.094; \quad   a^2_0=-0.022\pm0.025; \quad   r=0.988
\end {equation}
The resulting subtraction terms $\lambda^I_0(s)$ with fitting
straight lines~(\ref{f3}) are shown in Fig.2.\\
The uncertainties of the $\pi\pi$ lengths $a^0_0$ and $a^2_0$ are
defined both by the statistical errors of values
$\lambda^I_0(s_i)$, which are expressed through the uncertainties
phase shifts by means of the standard rule of propagation of errors
and by the theoretical uncertainties $\Delta\varphi^I_0(s)$.\\
It may seem that the lengths obtained from the isotopic channel I=2 are
in accordance with the GChPT version, but it is not so.
The large uncertainties of the obtained lengths do not permit to
make any choice at all using this channel only.
On the other hand, obviously, main information about the $\pi\pi$
lengths is contained in the term $\lambda^0_0$, because it
concentrates in itself the overwhelming part of general
statistics.
We discuss the obtained solutions of the Roy equations for each
isotopic channel in detail in order to show that these solutions
((\ref{f6}~and~(\ref{f8})) are in accordance with each other
within the error limits and to demonstrate that the subtraction
terms $\lambda^I_0(s)$ are really  linear functions of s.
For us, it is an additional proof that all the calculation steps in
the solving the Roy equations and also all preliminary work
comprising the fitting the phase shifts $\delta^I_l$ were carried out
correctly.\\
The final result was obtained on the basis of the use of all
available statistics, i.e., by both isotopic channels, I=0 and I=2:
\begin {equation}
 \label{f9}
         a^0_0=0.212\pm0.015; \quad   a^2_0=-0.043\pm0.010 
\end {equation}
with the correlation coefficient {\bf r}=0.945.
Now we can make a preliminary conclusion: the obtained results
unambiguously witness in favor of the standard ChPT version and
exclude the GChPT one.
More detailed discussion of the obtained results will be provided
in section~4.\
In conclusion of the present section we shall demonstrate the
results obtained by using the phase shifts from~\cite{Hoog}.
If the $S_2$-wave phase shifts are used only from~\cite{Hoog} then
the following is obtained:\\
 a) $\delta^2_0$ -- Hoogland~\cite{Hoog} only
\begin {gather}
 I=0: \  a^0_0=0.209\pm0.015; \quad   a^2_0=-0.036\pm0.011; \
  r=0.978; \  \chi^2/NDF=123/106 \label{f10}\\
 I=2: \  a^0_0=0.140\pm0.071; \quad   a^2_0=-0.079\pm0.016; \
  r=0.986; \quad   \chi^2/NDF=32.5/5 \label{f11}
\end {gather}
It is obvious that the results obtained in the different isotopic
channels are contradictory for the parameter $a^2_0$.
Moreover, a linearity test is not satisfied - the value of $\chi^2$
 in the channel I=2 (\ref{f11}) shows that the subtraction term $\lambda^2_0(s)$,
obtained by using the phase shifts $\delta^2_0$ from~\cite{Hoog},
is not a linear function of s.\\
When the both isotopic channels are used then the following is
obtained:
\begin {equation}
 \label{f12}
  a^0_0=0.163\pm0.015; \quad   a^2_0=-0.071\pm0.009; \   r=0.904
\end {equation}
Thus, when the phase shifts $\delta^2_0$ from~\cite{Hoog} were used
 a concordance was absent both between the results obtained in
the different isotopic channels as well as with the
solution~(\ref{f9}), obtained by using the rest of the phase shifts
$\delta^2_0$ data base, taken from [18-22].\\
The using of the united phase shifts $\delta^2_0$ data base from
[18-23] does not change the situation in principle by force of
statistical domination of the phase shifts from~\cite{Hoog}.\\
 b) $\delta^2_0$ -- Hoogland~\cite{Hoog} + all the rest
\begin {gather}
 I=0: \  a^0_0=0.209\pm0.015; \quad   a^2_0=-0.039\pm0.011; \
  r=0.979; \  \chi^2/NDF=125/106  \label{f13}\\
 I=2: \  a^0_0=0.182\pm0.075; \quad   a^2_0=-0.064\pm0.018; \
  r=0.992; \quad  \chi^2/NDF=92/32  \label{f14}
\end {gather}
 It was got by using both channels:
\begin {equation}
 \label{f15}
  a^0_0=0.177\pm0.015; \quad   a^2_0=-0.064\pm0.010; \   r=0.921
\end {equation}
It should be noted that adding the phase shifts $\delta^2_0$
from~\cite{Hoog} in the channel I=0, which is the main source of
the information about $a^0_0$ and $a^2_0$, leads to the systematic
increase of the value $a^2_0$ (13).
%
\section{Correlation  between  $\delta^0_0(s)$ and $\delta^2_0(s)$}
%
In the previous section it was shown that the strongly correlated S-wave $\pi\pi$ lengths
in the result of solution of the Roy equations  were obtained.
Such correlation, with $r\,\cong1$, implies that the values $a^0_0 $ and $a^2_0 $ are related
by a linear dependence.
But this fact signifies that the phase shifts $\delta^0_0(s)$ and $\delta^2_0(s)$
by force of the near threshold expansion $ \delta^I_0 (s) \propto a^I_0\,q $ must be related
by a linear dependence too in some energy region near the threshold.
We do not know only the range of this region .
We analyzed the ratio $\xi(s)=\delta^0_0(s)/\delta^2_0(s)$  for the available
experimental data to study this problem.
No evident dependence on s was found   in the behavior of $\xi(s)$  from the threshold
up to s=42, i.e., up to $m_{\pi\pi}$=900MeV (Fig.3).\\
In such a way, the simplest 0-hypothesis to verify  is the hypothesis about
proportionality phase shifts in some area above the threshold.
As the phase shifts $\delta^0_0(s_i)$   and $\delta^2_0(s_j)$ were
measured mainly at different  energy values, the smoothed curve (Fig.1) representing
the fitting function~({4})  was used for calculation of the phase shifts
$\delta^2_0$ at the points $s=s_i$, where the phase shifts $\delta^0_0$  were measured.
Thus, the ratio of the S-wave phase shifts was calculated as
$\xi (s_i) = \delta^0_0 (s_i)/\delta^2_0 (s=s_i)$.\\
The uncertainties  $ \sigma _ {\xi} $ were calculated  by the standard  rule of  propagation
of errors and finally they were defined both by errors of phase shifts $\delta^0_0(s_i)$
and  $\delta^2_0(s_j)$.
It was calculated by fitting   $\xi(s)\equiv\eta$--const,  for interval  $s=10\div42$:
$\eta=-4.66\pm0.05$;\  $\chi^2$/NDF=78/82.
The fitting $\xi (s) \equiv\eta$--const  for the interval $s=4\div42$ gives naturally
the same value of $ \eta$, because the statistical weights of the points near the threshold
are insignificant.
In general, large uncertainties  of the values $ \xi (s) $ near the threshold (Fig.3)
are caused by the fact that the phase shifts  $ \delta^0_0 (s) $ and $\delta^2_0 (s)$
have large relative errors in that region.\\
So, the proposed 0-hypothesis is confirmed by means of the statistical proof.
And consequently  we  can conclude that within the present accuracy of the experimental
data the ratio  of  S-wave phase  shifts does not depend on the energy in the
wide  enough region.
Thus, for  this  energy  region, i.e., for  $s=10\div42$,    the relation    take  place:
\begin {equation}
 \label{f16}
               \delta^0_0(s)=\eta\,\delta^2_0(s)  
\end {equation}
\noindent where $ \eta =-4.65\pm0.05 $.
So, as stated above, from the fact of strong correlation of the $\pi\pi$ lengths
follows  linear dependence  of the phase  shifts near the threshold .
Then we found the proportionality between $\delta^0_0(s)$ and $\delta^2_0(s)$
in some region above the threshold:  $s=10\div42$.
Our only proposal based on these facts  is that we  deal with the same proportionality.
I.e., we believe that the relation (16) is true from the threshold up to s=42.
Hence, in force by the near threshold expansion $ \delta^I_0 (s) \propto a^I_0\,q $,
the new constraint on scattering lengths follows:
\begin {equation}
 \label{f17}
                    a^0_0=\eta\,a^2_0   
\end {equation}
So, an  opportunity  appears  using the constraint (17) to eliminate the correlation
between $a^0_0 $ and $a^2_0 $ in the process of the  subtraction terms  fitting.
As the result, the solution  was obtained, which  we denote as "$\eta$-solution":
\begin {equation}
 \label{f18}
         a^0_0=0.211\pm0.005; \quad   a^2_0=-0.0454\pm0.0010 
\end {equation}
Here it should be emphasized that  $\eta$-solution is in accord with the solution (9)
obtained without using additional constraints.
It is a very important  point.
This signifies that the additional constraint (17) relating the $\pi\pi$ lengths
does not correct the Roy equations but  eliminates the  correlation only,
when  subtraction terms are fitting.\\
%
  \begin{figure}[t]
    \includegraphics[width=\textwidth, height=4.5in]{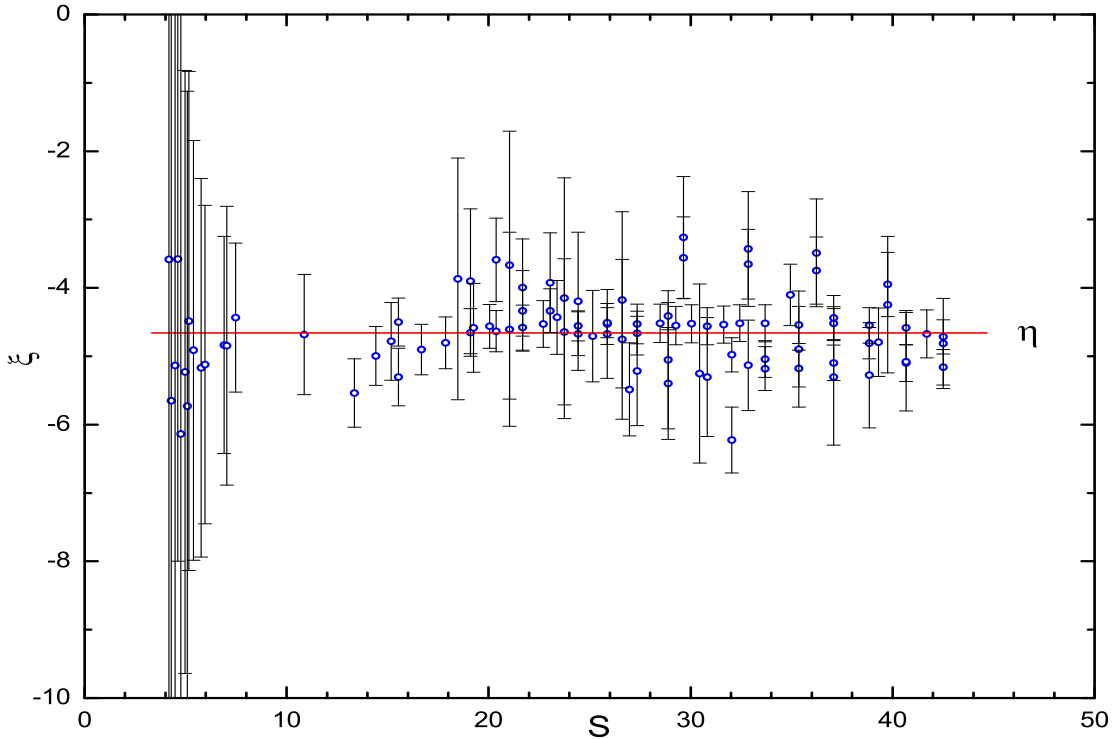}
    \caption{The ratio of the S-wave phase shifts  $\xi(s)=\delta^0_0(s)/\delta^2_0(s).$
                  The straight  line   represents   the   constant  $\eta=-4.66.$}
  \end{figure}
%
In this sense the condition (17) is a new independent constraint on S-wave $\pi\pi$ lengths.
We stress than the process of obtaining the $\eta$-solution and  the solution (9) is the same
right up to calculating subtraction terms $\lambda^I_0(s)$ inclusive.
The difference between them consists in using the constrain (17)  for obtaining
the $\eta$-solution on the fitting step.
The solution (9) was obtained without using of any additional constraints.
The obtained results are presented in Fig.6 and Fig.7.
%
\section{Discussion  and  Summary}
%
Let us analyse obtained results in more detail.
We should start  by  comparison of our result~(9) with the theoretical prediction  received
in~\cite {Col1}, in which ChPT calculations were supplemented with the phenomenological
 representations based on the Roy equations~\cite {Anan}:
\begin {equation}
 \label{f19}
         a^0_0=0.220\pm0.005; \quad   a^2_0=-0.0444\pm0.0010  
\end {equation}
These results are in good accord with each other for both parameters $a^I_0 $
within error limits.
Hence  our result~(9) certainly witnesses in favor of the standard ChPT version and
excludes GChPT one, with $a^0_0=0.263 $.
Thus, the problem  of choosing the true ChPT version, in our opinion, is solved. \\
But it is possible to put a more tough  question:
whether  there is statistically significant conformity between the theoretical result~(19) and
the result of the model--independent
\clearpage
%
  \begin{figure}[t]
    \includegraphics[width=\textwidth, height=4.0in]{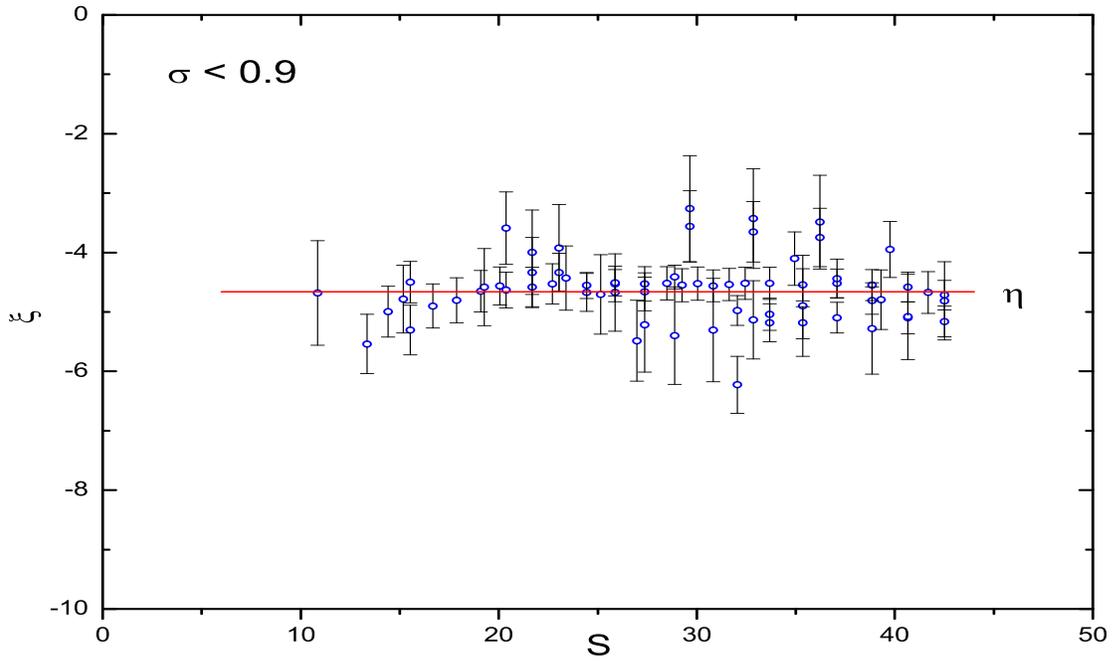}
    \caption{The ratio of the S-wave phase shifts  $\xi(s)=\delta^0_0(s)/\delta^2_0(s)$
                  after the filtration, $\sigma_{\xi} < 0.9$.
                  The straight  line   represents   the   constant  $\eta=-4.66.$}
  \end{figure}
%
  \begin{figure}[t]
    \includegraphics[width=\textwidth, height=4.0in]{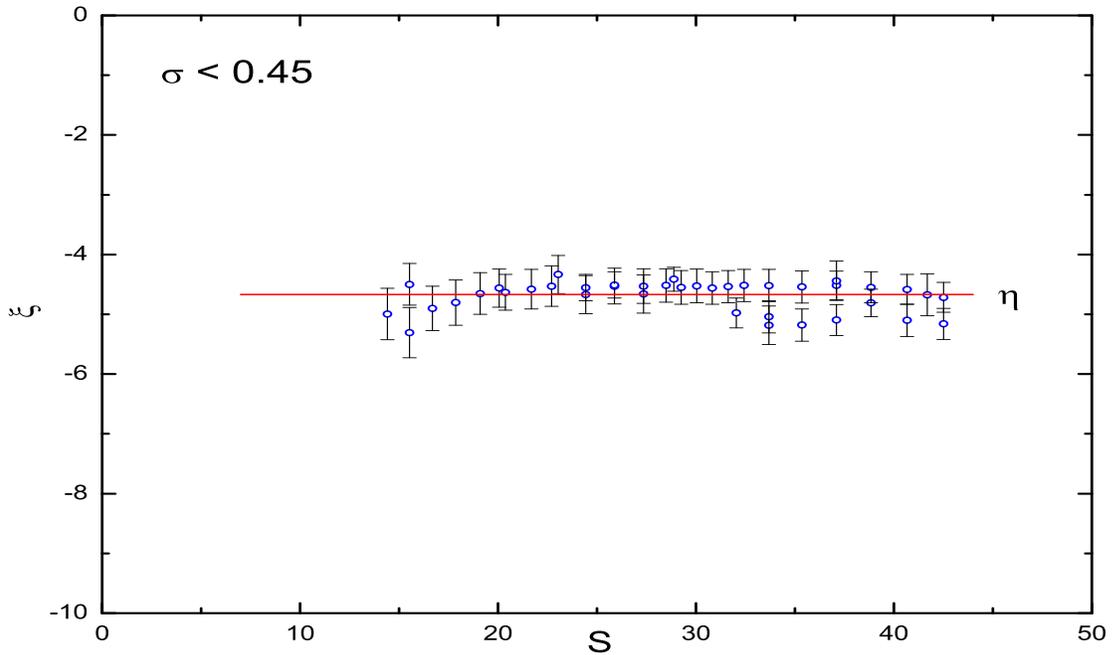}
    \caption{The ratio of the S-wave phase shifts  $\xi(s)=\delta^0_0(s)/\delta^2_0(s)$
                  after the filtration, \qquad  \qquad $\sigma_{\xi} < 0.45$.\
                  The straight  line   represents   the   constant  $\eta=-4.66.$}
  \end{figure}
%
\clearpage
\noindent
 analysis~(9)?
It is seen that $1\sigma $ contour  ellipses do not intersect (Fig.6). Is  the hypothesis
true that these results are statistically consistent one with the other or may be
there is  a significant statistical discrepancy of these results? \\
Here it is necessary to take into account one feature which  distinguish comparison
of the results in an one-dimensional case and in a plane.
In a plane the probability of a random variable to get inside a $ 1\sigma $
contour  ellipse is equal  P=0.39.
Certainly it is not enough  to draw final conclusions.
Therefore it is more correct to compare $ 2\sigma $ contour  ellipses.
The probability to get in such an ellipse is equal  P=0.865. The ChPT solution~(19) gets in
the border of the $ 2\sigma $  contour  ellipse~(9) (Fig.6). I.e., we have to reject the
hypothesis that the results~(9) and~(19) are consistent  with probability 13.5\%.
This probability  is very large.\\
All this taken together forces us to come to a conclusion, that we do not have sufficient
base to reject a hypothesis about the statistical agreement of the results~(9) and~(19).
Thus, we come to the conclusion, that the solutions~(9) and~(19) are statistically consistent and
do not contradict each other. \\
Let us carry out comparison with other works in which the results of experiment $K_{e4}$ E865
for calculation of S-wave $\pi\pi$ lengths  were used.
In the work ~\cite {Pis}, where the final results of this experiment were presented,
it was received  without using of the additional relations linking  $a^0_0$ and $a^2_0$:
\begin {equation}
 \label{f20}
   a^0_0=0.203\pm0.033\pm0.004_{syst}; \quad  a^2_0=-0.055\pm0.023\pm0.003_{syst}
\end {equation}
I.e., we have full conformity with our result~(9) within the limits of
errors (Fig. 7).
In the work~\cite {Anan} the position and the borders of the area in the plane
($a^0_0, a^2_0 $) in which S-wave lengths are consistent with  the Roy equations solution
and the available experimental data on  $\pi\pi$ phase shifts above 0.8 GeV were specified.
It was received for the central curve of this area:
\begin {equation}
 \label{f21}
   a^2_0=-0.0849+0.232a^0_0-0.0865(a^0_0)^2 \ [\pm0.0088]
\end {equation}
The value given in brackets defines the  width of the band. In Fig.6 and Fig.7 this band is
designated as UB (universal band). \\
In the work~\cite {Des} the calculations done in~\cite {Anan} were repeated with some changes
and  practically the same parameters describing UB were received. Further, using the obtained
parametrization and the experimental data including the data~\cite {Pis}, the authors  received:
\begin {equation}
 \label{f22}
              a^0_0=0.228\pm0.013; \quad  a^2_0=-0.0380\pm0.0044    
 \end {equation}
with the factor of correlation {\bf r}=0.799.
In the work~\cite {Pis}  using  UB~\cite {Anan} as the  additional constraint close
results were received.\\
The results obtained in~\cite {Pis} and~\cite {Des} are given in Fig.7.
The solution~(22) gets in our $2\sigma $  contour ellipse as well as our solution~(9)
gets in $2\sigma$  contour ellipse of the solution~(22). Thus, it is possible
to state that the results~(9) and~(22) do not contradict  one another. \\
Let us consider the problem of stability of the received solution~(9) concerning the procedure
of experimental data selection, i.e.,  $\pi\pi$ phase shifts, which in our method of the
solution of the Roy equations  are utilized  as input.
Stability of the solutions  versus  variations of the initial data is an important indicator
of reliability of the method of the solution and consistency of the initial data.
We have shown above  that the use of the data from~\cite {Hoog} leads to contradictious results.
Further, the results of an expanded analysis are presented.
\newpage
%
  \begin{figure}[!t]
    \includegraphics[width=\textwidth, height=4.5in]{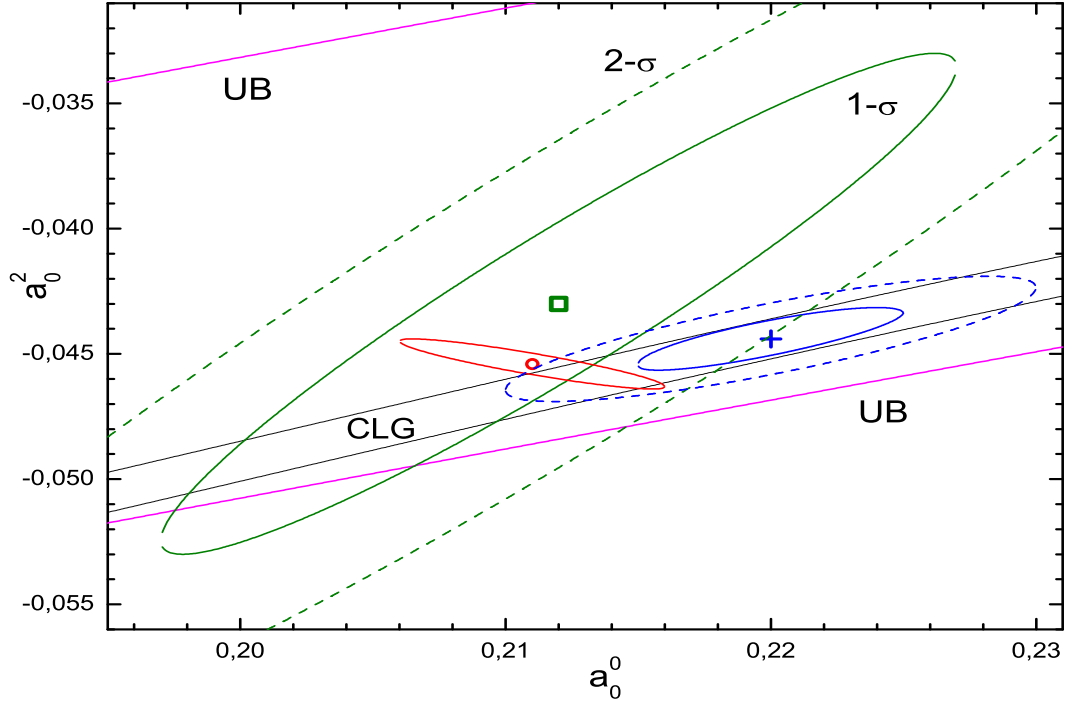}
    \caption{(Color online) The  S-wave $\pi\pi$  lengths. The olive ellipse with the centre
       as a square indicates the solution~(9); the solid line - $1\sigma$  ellipse,
       the dashed one - $2\sigma$.
       The red ellipse with the centre as a circle represents  $\eta$-solution~(18).
       The blue ellipse with the centre as a cross shows the  ChPT result~\cite {Col1};
       the solid line - $1\sigma$  ellipse, the dashed one - $2\sigma$.
       The straight lines marked UB indicates the  area allowed for S-wave lengths~\cite {Anan}.
       The strip marked CLG is the range   corresponding the chiral constraint~\cite {Col2}.}
 \end{figure}
\noindent
1) Change of the data sets  used.\\[1mm]
a) The solution  of the Roy equations  without the  phase shifts  $ \delta^2_0 $
which  calculated on the basis of the cross sections received in~\cite {Kerm}
(Tabl. 1):
 \begin {equation}
 \label{f23}
      a^0_0=0.213\pm0.015; \quad   a^2_0=-0.044\pm0.011;  \quad  r=0.961  
\end {equation}
b) Calculation of S-wave lengths without the data from  the work~\cite {Alek}:
\begin {equation}
 \label{f24}
       a^0_0=0.208\pm0.015; \quad   a^2_0=-0.045\pm0.010; \quad  r=0.944 
 \end {equation}
2) Change of the degree of the fitting S-wave phase shifts  polynomial~(4): \\
For  m=5  in the formula~(4) we  receive:
\begin {equation}
 \label{f25}
       a^0_0=0.211\pm0.013; \quad   a^2_0=-0.044\pm0.009;  \quad  r=0.918  
\end {equation}
Comparison of the results (23-25) with the above  solution~(9) shows, that the criterion of
the stability for the given solution is satisfied. \\
Let us proceed to the discussion of observable proportionality of S-wave phase shifts.
It may seem, that the values $\xi (s_i)$ in  Fig.3 have a wide scatter  and, therefore,
can be described not only by a constant, but also by some class of smooth functions of s.
But these doubts are
%
  \begin{figure}[!t]
    \includegraphics[width=\textwidth, height=4.5in]{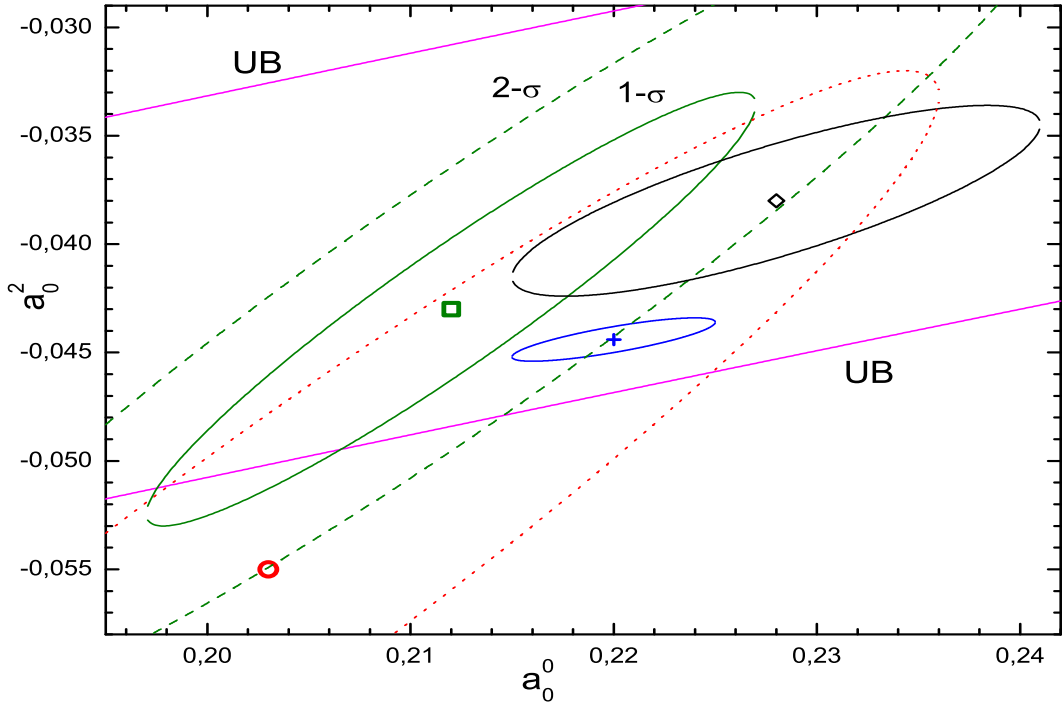}
    \caption{(Color online) The S-wave $\pi\pi$  lengths. The olive ellipse with the centre
         as a square  indicates the solution~(9); the solid line - $1\sigma$  ellipse,
         the dashed one - $2\sigma$.
         The black ellipse with the centre as a rhomb represents the  result~\cite {Des}.
         The red dotted ellipse with the centre as a circle represents the  result~\cite {Pis}.
         The blue ellipse with the centre as a cross  shows  the  ChPT  result~\cite {Col1}.}
\end{figure}
\noindent
based on visual illusion.
The point is that the values $\xi (s_i)$ with the large errors $ {\sigma_{\xi}}$
form "a cloud" which masks true dependence.
These points have small statistical weights and do not give  contribution
to  $\eta$ value.
We  have carried out a filtration leaving only  the points with
errors less than given, i.e., with $ \sigma _ {\xi} < \sigma_k $ where $ \sigma_k $
lay in the range $1\div0.3 $.
The remaining after the filtration values of  $\xi (s_i)$ were fitted by a constant.
The result: all received  $\eta_k$ lay in the range $ (-4.66\div-4.68) $ and have
errors $ \sigma _ {\eta} =0.05 $ and goodness-of-fit test is satisfied:
$ \chi^2 < N_k $, where $N_k $ - number of points $\xi (s_i)$ after the filtration
with parameter $ \sigma_k $.
I.e., all sets of points $\xi_i$ after the filtration are well described by a constant.
The results are given in Fig. 3-5.
Fig. 3 - the values $\xi (s_i)$ without the filtration, \ Fig.4 - $ \sigma_k=0.9$,
\ Fig.5 - $ \sigma_k=0.45$. \\
One can easily see  that after rejection of the points with  large errors, the remaining
points more and more concentrate near the straight line.
Thus, if the 0-hypothesis is that values $\xi (s_i)$ are the constant within the
considered region, this hypothesis are proved both statistically and visually. \\
Additionally, a linear function was utilized for fitting $\xi (s_i)$ also  to find any
dependence of $\xi$ on s, if it  exists nevertheless.
The linear function may be represented as  $f(s = \eta'+ b\,(s-4)$,
where $\eta'$ and b are free parameters.
It was obtained as the   result of fitting for  $s=10\div42$:
 \begin {equation}
 \label{f26}
         \eta'=-4.51\pm0.15; \quad   b=-0.006\pm0.006   
\end {equation}
 $\chi^2$/NDF=77/81.\ As $ |b | \le \sigma _ {b}$, there is no reason to believe,
that the  hypothesis about the linear dependence $\xi (s)$ is confirmed. \\
Let us consider the problem of  influence of the $\eta$ errors   on  the $\pi\pi$ lengths errors
obtained by the solving of the Roy equations  ($ \eta $-solution~(18)).
It may seem that  the  sizes of $ a^I_0$ errors  are small because  the   sizes
$ \sigma_{\eta}$  are small. But it is not so.
The basic contribution to errors of the values $a^0_0 $ and $a^2_0 $, in this case,
is brought by  theoretical uncertainties in the Roy equations $ \Delta\varphi^I_0 (s)$.
Role of the additional constraint~(17) is only to eliminate the  correlation between
$a^0_0 $ and $a^2_0 $ in the process of fitting of subtraction terms $ \lambda^I_0 (s)$.
To show it we  increase $ \sigma _ {\eta}$  four  times,
i.e., we used $ \sigma _ {\eta} =0.2 $. In result the following $\eta$-solution is obtained:
 \begin {equation}
 \label{f27}
         a^0_0=0.211\pm0.0052; \quad   a^2_0=-0.0454\pm0.0016   
 \end {equation}
The solution~(27)  shows a weak dependence on $\sigma_{\eta}$. \\
In the works~\cite {Col2, Col3} it was shown that the width of the allowed area in
the  ($a^0_0 $,\ $a^2_0 $) plane  can be reduced considerably  by  using the additional chiral
constraint  imposed on S-wave lengths.
This constraint links the combination ($2a^0_0 - 5a^2_0 $) with the  scalar pion radius
$ <r^2_s> $. In the result of utilizing of this constraint it was received:
\begin {equation}
 \label{f28}
 \Delta a^2_0=0.236\Delta a^0_0-0.61(\Delta a^0_0)^2-9.9(\Delta a^0_0)^3 \ [\pm0.0008]
 \end {equation}
where $ \Delta a^0_0=a^0_0-0.22 $; \ $ \Delta a^2_0=a^2_0+0.0444.$
In Fig.6 this narrow strip is denoted as CLG. From this figure it follows that
$\eta$-solution~(18) lay  practically in the  border of the CLG-band  and half of
$ 1\sigma $ contour  ellipse overlaps  this band.
Also $\eta$-solution lays practically in the  border of  $2\sigma$  contour  ellipse of the
ChPT-solution~(19).
Owing to all aforementioned, one may conclude  that $\eta$-solution~(18)
received under the additional condition~(17) is consistent both with the  chiral
CLG constraint~(28) within the  $1\sigma$  level and with  the ChPT-solution~(19) within the
$2\sigma $ level . \\
It is natural to compare the value of  $\eta =-4.66\pm0.05 $ received in the present work
with $\eta_{ChPT}=a^0_0/a^2_0$, which follows from the chiral theory.
Only it is necessary to take into account that the calculation of the ratio of S-wave $\pi\pi$
lengths should be  carried  out in view of their  correlation.
From values $a^0_0 $, $a^2_0 $ and $2a^0_0-5a^2_0 $ obtained  within ChPT
framework~\cite {Col1, Col3}, one may estimate the factor of correlation between
$a^0_0 $ and $a^2_0 $. It is equal 0.8.
In view of it for the  S-wave $\pi\pi$ lengths ratio it was  received
$ \eta _ {ChPT} =-4.95\pm0.21 $. The difference from the $\eta$ value   received by us
is slightly more than one $ \sigma $. \\
Summarizing the main results of the present study, one may say
that the solutions received by us and other authors~\cite {Pis, Des} (Fig.7) are grouped
near the ChPT-solution~\cite {Col1} and are consistent both with each other and with this
ChPT solution.
Thus, we believe that the problem of  choosing of the scenario of chiral symmetry violation
is solved.
The available mismatch on the $1\sigma $  level both among the considered  Roy equations
solutions and among these solutions and the theoretical
prediction~\cite {Col1} may be caused by the fact that we used non identical
sets of the experimental data and different methods of  the Roy equations solution.
Therefore, it seems  that prior to search for the physical reasons of such divergence,
it is necessary to come to an agreement about using of uniform experimental data base.
Also it is desirable to organize the procedure  of the Roy equations solution in
such a way that enables to check both individual solutions in every
isotopic  channel and monitor influence of various errors
(statistical, systematic, theoretical, errors from additional constraints)
on the resulting errors of the S-wave $\pi\pi$ lengths.
May be that such unification of the initial data and more detailed  control
of the course of the solution will allow to reduce the existing
discrepancy.\\[3mm]
This work was supported in part by the Russian Foundation for
Basic Research (project no. 00-02-17852).\\[6mm]
%
{\large Appendix A}\\
\begin{center}
{\large Table 1.}\\[8pt]
\begin{tabular}{|c|c|c|c|}\hline
\large \quad s \quad \rule{0pt}{13pt}&\ E, MeV\ \ &\ $\ \delta^2_0$, deg\ \ &
 $\ \ \sigma (\delta^2_0)$, deg\ \ \\[3pt]\hline
4.15 & 284.3 & -1.11 & 0.67 \rule{0pt}{13pt} \\\hline
4.45 & 294.3 & -1.75 & 0.96 \rule{0pt}{13pt} \\\hline
4.75 & 304.2 & -1.05 & 1.11 \rule{0pt}{13pt} \\\hline
5.05 & 313.7 & -0.71 & 1.26 \rule{0pt}{13pt} \\\hline
5.35 & 322.8 &  0.93 & 2.30 \rule{0pt}{13pt} \\\hline
5.65 & 331.8 & -5.20 & 3.27 \rule{0pt}{13pt} \\\hline
\end{tabular}\\[8mm]
\end{center}
{\large Appendix B}\\ \\
$ K^0_{10}=\frac{s-4}{(x-s)(x-4)}+\frac2{3x} \left[\frac{x}{s-4} ln \left(
 \frac{x+s-4}x \right)-1 \right]-\frac{2(s-4)}{3x(x-4)} $ \\[3mm]
$ K^0_{20}=\frac3x \left\{2\left(1+\frac{2s}{x-4}\right) \left[\frac{x}{s-4}
  ln\left(\frac{x+s-4}x \right)-1\right] + \frac{s-4}{x-4}\right\} $ \\[3mm]
$ K^0_{30}=\frac5{3x}\left\{ 2\left[\frac{x}{s-4} ln\left(\frac{x+s-4}x\right)-1\right] +
 \frac{s-4}{x-4}\right\} $ \\[5mm]
$ K^1_{11}=\frac13\left\{ \frac4{s-4}\left[ \left(\frac12 + \frac{x}{s-4} \right)
  ln\left(\frac{x+s-4}x\right)-1\right] - \frac{s-4}{3x(x-4)} \right\}$ \\[3mm]
$ K^1_{21}=\frac{s-4}{(x-s)(x-4)}+\frac6{s-4} \left(1+\frac{2s}{x-4} \right)
  \left[ \left(\frac12 + \frac{x}{s-4} \right)ln\left(\frac{x+s-4}x\right)-1\right] -
  \frac{3(s-4)}{2x(x-4)} $ \\[3mm]
$ K^1_{31}=-\frac53\left\{ \frac2{s-4}\left[ \left(\frac12 + \frac{x}{s-4} \right)
  ln\left(\frac{x+s-4}x\right)-1\right] -  \frac{s-4}{6x(x-4)}\right\} $ \\[5mm]
$ K^2_{10}=\frac1{3x}\left\{ 2\left[\frac{x}{s-4} ln\left(\frac{x+s-4}x\right)-1\right] -
 \frac{s-4}{x-4}\right\} $ \\[3mm]
$ K^2_{20}=-\frac3x \left\{\left(1+\frac{2s}{x-4}\right) \left[\frac{x}{s-4}
  ln\left(\frac{x+s-4}x \right)-1\right] - \frac{s-4}{2(x-4)}\right\} $ \\[3mm]
$ K^2_{30}=\frac{s-4}{(x-s)(x-4)}+\frac1{3x} \left[\frac{x}{s-4} ln \left(
 \frac{x+s-4}x \right)-1 \right]-\frac{5(s-4)}{6x(x-4)} $
%


\begin{thebibliography}{99}
\bibitem{Gas} J.Gasser and H.Leutwyler, Phys. Lett. B125 (1983)321; 325
\bibitem{Bij} J.Bijnens et al., Phys. Lett. B374(1996)210
\bibitem{Kne} M.Knecht et al., Nucl. Phys. B457 (1995) 513; B471(1996)445
\bibitem{Ros} L.Rosselet et al., Phys. Rev. D15 (1977)574
\bibitem{Pat} O.O.Patarakin, V.N.Tikhonov, K.N.Mukhin, Nucl. Phys. A598(1996)335
\bibitem{Mai} V.N.Ma\u{\i}orov, O.O.Patarakin, V.N.Tikhonov, Yad. Fiz. 63(2000)1699
              [Phys. At. Nucl. 63 (2000) 1612]
\bibitem{Mai2}V.N.Ma\u{\i}orov, O.O.Patarakin, Preprint IAE-6274/4 (Moscow, 2003); Hep-ph/0308162
\bibitem{Pis} S.Pislak et al., Phys. Rev. D67 (2003) 072004
\bibitem{Roy} S.M.Roy, Phys. Lett. B36  (1971) 353
\bibitem{Bas} J.L.Basdevant, C.D.Frogatt and  L.Petersen, Nucl. Phys. B72(1974)413
\bibitem{Pen} M.R.Pennington, S.D.Protopopescu, Phys. Rev. D7 (1973)1429
\bibitem{Pro} S.D.Protopopescu et al., Phys. Rev. D7 (1973) 1279
\bibitem{Eng} A.Engler et al., Phys. Rev. D10 (1974) 2070
\bibitem{Alek}E.A.Alekseeva et al., Zh. Eksp. Teor. Fiz. 82(1082)1007
              [Sov. Phys. JETP 55(1982)591]
\bibitem{Est} E.A.Estabrooks  and   A.D.Martin, Nucl. Phys. B79 (1974)301
\bibitem{Car} J.Carroll et al., Phys. Rev. D10(1974)1430
\bibitem{Kam} R.Kaminski et al., Z. Phys. C74(1997)79
\bibitem{Lost}J.Losty et al., Nucl. Phys. B69 (1974)185
\bibitem{Colt}E.Colton et al., Phys. Rev. D3 (1971)2028
\bibitem{Coh} D.Cohen et al., Phys. Rev. D7 (1973) 661
\bibitem{Dur} N.B.Durusoy et al., Phys. Lett. B45 (1973)517
\bibitem{Bek} G.V.Beketov et al. Yad. Fiz. 19 (1974) 1032
              [Sov. J. Nucl. Phys. 19 (1974) 528]
\bibitem{Hoog}W.Hoogland et al., Nucl. Phys. B126 (1977) 109
\bibitem{Kerm}M.Kermani, O.Patarakin, G.R.Smith et al., Phys. Rev. C58 (1998)3431
\bibitem{Anan}B.Ananthanarayan et al., Phys. Rep. 353 (2001) 207; Hep-ph/0005297
\bibitem{Des} S.Descotes et al., Eur. Phys. J. C24 (2002) 469; Hep-ph/0112088
\bibitem{Col1} G.Colangelo, J.Gasser and H.Leutwyler, Phys. Lett. B488(2000)261;
               Hep-ph/0007112
\bibitem{Col2} G.Colangelo, J.Gasser and H.Leutwyler, Phys. Rev. Lett. 86(2001)5008;\\
               Hep-ph/0103063
\bibitem{Col3} G.Colangelo, J.Gasser  and  H.Leutwyler, Nucl. Phys. B603 (2001)125;
               Hep-ph/0103088
\end{thebibliography}
\end{document}